\documentclass[preprint2]{aastex}

\newcommand{\twcott}{$^{12}$CO $J$=3$-$2}
\newcommand{\twcoto}{$^{12}$CO $J$=2$-$1}

\newcommand{\twco}{$^{12}$CO}

\newcommand{\jtt}{$J$=3$-$2}
\newcommand{\jto}{$J$=2$-$1}
\newcommand{\joz}{$J$=1$-$0}

\newcommand{\etal}{et al.}
\newcommand{\eg}{e.g.}
\newcommand{\ie}{i.e.}
\newcommand{\h}{$^{\rm h}$}
\newcommand{\m}{$^{\rm m}$}

\newcommand{\kms}{km s$^{-1}$}
\newcommand{\tas}{$T_{\rm A}^{\star}$}

\newcommand{\tmb}{$T_{\rm MB}$}
\newcommand{\msun}{M$_\odot$}

\shorttitle{CO in Double Barred Galaxies}
\shortauthors{Petitpas \& Wilson}

\begin{document}

\title{Molecular Gas in Candidate Double Barred Galaxies III. A Lack of Molecular Gas?}
\author{Glen R. Petitpas}
\affil{University of Maryland}
\affil{College Park MD, USA 20742}
\email{petitpas@astro.umd.edu}
\and
\author{Christine D. Wilson}
\affil{McMaster University}
\affil{1280 Main Street West, Hamilton ON,
Canada L8S 4M1}
\email{wilson@physics.mcmaster.ca}

\begin{abstract}

Most models of double-barred galaxies suggest that a molecular gas
component is crucial for maintaining long-lived nuclear bars.  We have
undertaken a CO survey in an attempt to determine the gas content of
these systems and to locate double barred galaxies with strong CO
emission that could be candidates for high resolution mapping. We
observed ten galaxies in CO \jto\ and \jtt\ and did not detect any
galaxies that had not already been detected in previous CO
surveys. We preferentially detect emission from galaxies containing
some form of nuclear activity. Simulations of these galaxies require
that they contain 2 to 10\% gas by mass in order to maintain
long-lived nuclear bars. The fluxes for the galaxies for which we have
detections suggest that the gas mass fraction is in agreement with
these models requirements. The lack of emission in the other galaxies
suggests that they contain as little as $7 \times 10^6$ \msun\ of
molecular material which corresponds to $\lesssim$0.1\% gas by
mass. This result combined with the wide variety of CO distributions
observed in double barred galaxies suggests the need for models 
of double-barred galaxies that
do not require  a large, well ordered molecular gas component.

\end{abstract}

\keywords{Galaxies: nuclei --- galaxies: active}

\section{Introduction \label{intro}}

Double-barred galaxies have been proposed as a means of transporting
molecular gas interior to inner Lindblad resonances (ILRs) where it
may fuel starbursts or other forms of nuclear activity \citep{shl89}.
Thus far, double-barred galaxies have been identified predominantly
through the analysis of near infrared (NIR) images
\citep[\eg][]{mul97} and manifest themselves as variations in the
position angles and ellipticity with galactic radius. A variety of
models have been proposed to explain the nature of these nuclear bars,
with origins varying from kinematically distinct nuclear bars to
nuclear agglomerations that co-rotate at the same speed as the large
scale bar \citep{fri93,sha93}. To allow long-lived nuclear bars, these
models usually require substantial amounts of dissipative gas to
prevent the nuclear stellar populations from suffering rapid kinematic
heating and subsequent bar destruction.

There is another class of models that attempt to explain nuclear bars
using purely stellar orbits. It is known that there are different
classes of orbits in a barred potential.  The two important ones are
the $x_1$ family that runs parallel to the bar major axis, and the
$x_2$ family that runs perpendicular to it \citep[\eg,][]{ath92}. It
was thought that the $x_2$ orbits of the large scale bar near the
nucleus could form the $x_1$ orbits of the smaller nuclear bar and the
corotation radius of the nuclear bar could correspond to the ILR of
the large scale bar. In this picture, the nuclear bar must {\it
always} be aligned perpendicular to the main bar. \citet{fri93} rule
out this model by studying a large sample of double-barred galaxies,
since they found that not all the observed offset angles between the
nuclear bar and the main bar can be explained by inclination
effects. Recently, \citet{mac00} find that there exist orbits in which
particles in a double-barred potential (where the inner bar rotates
faster than the large scale bar) remain on closed orbits and may form
the building blocks of long-lived double-barred galaxies without the
need for a gaseous component.

\defcitealias{pet02a}{Paper~I}
\defcitealias{pet03}{Paper~II}

High resolution observations of the dynamics of these galaxies will
allow us to test these competing models and learn the true nature of
nuclear bars. Of the currently known double-barred galaxies, only a
few have been studied in detail using high resolution observations of
the molecular gas \citep[\eg][]{pet02a,jog99}. In \citeauthor{pet02a}
(\citeyear{pet02a}; hereafter \citetalias{pet02a}) we used high
resolution CO observations to compare the molecular gas distributions
of two candidate double-barred galaxies to the models of \citet{fri93}
and \citet{sha93}. We found that in NGC 2273 the molecular gas
emission takes the appearance of a bar-like feature that is aligned
with the NIR isophote twists. In NGC 5728, we observed a rather
disorderly molecular gas morphology that did not align with the NIR
morphology, nor did it align with any features seen at other
wavelengths in the nuclei of this galaxy.

The similarity in the NIR images of these galaxies suggests that the
galactic potentials may be similar. The variety of molecular gas
morphologies suggests that the molecular gas may have different
properties in each galaxy allowing it to respond differently to these
similar potentials. In \citeauthor{pet03} (\citeyear{pet03};
hereafter \citetalias{pet03}) we performed a multi-transition CO
survey of the nuclei of double barred galaxies for which high
resolution CO maps exist. We found that the molecular gas was cooler
(and less dense) in galaxies with more centrally concentrated gas
distributions and warmer and denser in galaxies with CO emission
scattered about the nucleus. The star formation rates in the galaxies
with non-centrally concentrated gas distributions tended to be higher
than in the galaxies with strong central concentrations. This result
suggests that either the gas distribution is influencing the star
formation activity, or that the star formation may be affecting the
gas properties.

The seven galaxies discussed in Papers I and II represent a small
fraction of the total number of galaxies known to have nuclear bars
(as indicated by NIR isophote twists). In order to strengthen the
hypotheses of those papers, we need to study a larger sample of
galaxies. Of the 93 galaxies studied to date
\citep{jar88,sha93,woz95,elm96,mul97} only 23 contain isophote twists
with size scales and position angle offsets large enough to be
resolved by the Caltech Millimeter and BIMA Arrays. Since the larger
NIR surveys mentioned above were performed using southern
observatories, most of the candidates are located in the southern
hemisphere. Of those 23 galaxies with resolvable bars, only 13 are at
a declination $> -$30$^{\circ}$. Six of these (NGC 470, NGC 2273, NGC
4736, NGC 5850, NGC 5728, and NGC 6951\footnote{More recent studies
suggest that NGC 6951 does not currently contain a nuclear bar, but
may have had one in the past \citep{per00}. See \S\ref{disc} for more
details.}) have high resolution CO maps
published \citep[\citetalias{pet02a};][]{jog98,won00,leo00,koh99}.
Only NGC 3945 was not included in our sample due to time and source
availability constraints.

We have performed a CO survey of the nuclei of 10 galaxies known to
have strong NIR isophote twists in an attempt to find CO-bright double
barred galaxies that would make good candidates for high resolution
mapping. Five of these galaxies (NGC 2273, NGC 3081, NGC 4736, NGC
5728, and NGC 6951) are discussed in detail in Papers I and II. In
\S\ref{obs} we discuss the observations and data reduction techniques.
In \S\ref{disc} we discuss our detections (and non-detections) in more
detail, and compare our observations to previous studies of these
galaxies. We also determine the molecular gas masses, and discuss the
implications of these masses to the double barred galaxy models. This
work is summarized in \S\ref{conc}.

\section{Observations and Data Reduction \label{obs}}

\subsection{NRAO Spectra}

The nuclei of nine double barred galaxies were observed in \twcoto\
using the National Radio Astronomy Observatory (NRAO)\footnote{The
National Radio Astronomy Observatory (NRAO) is a facility of the
National Science Foundation operated under cooperative agreement by
Associated Universities, Inc.} 12-m Telescope.  Observations were
taken in remote observing mode over a 14 hour period on 15 February,
2000. The half-power beam width of the NRAO 12-m was 29\arcsec\ at 230
GHz (\twcoto). All observations were taken in 2IF mode with the
Millimeter AutoCorrelator (MAC). The pointing was found to be accurate
to 6\arcsec\ for the first half of the evening when we observed our
galaxies with NGC $<$ 4736. This is poorer than the normal value for
the NRAO 12-m, likely due to the high winds. In the second half of the
evening the winds diminished, and the pointing improved to the more
normal value of 5\arcsec\ for observations of galaxies with NGC $\ge$
4736. The calibration was also monitored by observing spectral line
calibrators and planets and the spectral line calibrators agreed with
the published values. Thus, we adopt the nominal main beam efficiency
from the NRAO Users Guide of 0.29 at 230 GHz.

\subsection{JCMT Spectra}

Previous CO studies of double barred galaxies show CO \jtt/\jto\ line
ratios $\gtrsim 1$ \citepalias{pet03} so for galaxies that were not
detected with the NRAO Telescope, we obtained higher resolution
\twcoto\ and \twcott\ spectra using the James Clerk Maxwell Telescope
(JCMT)\footnote{The JCMT is operated by the Joint Astronomy Centre in
Hilo, Hawaii on behalf of the parent organizations Particle Physics
and Astronomy Research Council in the United Kingdom, the National
Research Council of Canada and The Netherlands Organization for
Scientific Research.}. These observations were taken over the period
of 1999 - 2000, mostly as part of a bad weather backup project. The
half-power beam width of the JCMT is 21\arcsec\ at 230 GHz (\twcoto)
and 14\arcsec\ at 345 GHz (\twcott).  All observations were obtained
using the Digital Autocorrelation Spectrometer. The calibration was
monitored by frequently observing spectral line calibrators. The
spectral line calibrators showed very little scatter from the
published values with individual measurements differing by typically
$< 15$\% from standard spectra. Thus, we adopt the nominal main beam
efficiencies from the JCMT Users Guide of 0.69 at 230/220 GHz and 0.63
at 345 GHz.  A detailed observing summary for the JCMT and NRAO
observations is given in Table \ref{obstable}.

\subsection{Reduction}

Similar data sets were averaged together using the software package
SPECX for the JCMT data and the Bell Labs data reduction package COMB
for the NRAO data. The data were binned to 10 km s$^{-1}$ resolution
(7.7 and 11.5 MHz at 230 and 345 GHz, respectively) and zeroth or
first order baselines were removed. The emitting regions we detected
were quite wide ($>$ 300 km s$^{-1}$) but the spectrometer bandwidth
was 800, 800, and 1200 km s$^{-1}$ for the NRAO and JCMT \jtt\ and
\jto data, respectively, which allowed for accurate baseline
determination. For the galaxies where we have no detections, the
baseline levels were set using the region of the spectrometer outside
a 400 \kms\ range centered on the rest velocity of the galaxies (\ie\
$V_{lsr} \pm 200$ \kms) in order to maximize our chances of detecting
any weak signal. The NRAO spectra for each galaxy are shown in Figure
\ref{nraospectra} and the JCMT spectra are shown in Figure
\ref{jcmtspectra}. In the cases where the NRAO spectra show no
convincing detections, we have included when available the \twcoto\
spectra taken at the JCMT and published in \citetalias{pet03} (Figure
\ref{morejcmtspectra}).  The spectral line intensities are summarized
in Table \ref{masstable}.

Using the $T_{\rm R}^*$ temperature scale would ensure that our
observed line strengths are as close to the true radiation
temperatures as possible \citep{kut81}. However, conversion to $T_{\rm
R}^*$ from $T_{\rm A}^*$ requires knowledge of the forward scattering
and spillover ($\eta_{\rm FSS}$), which is difficult to measure and
was not attempted during the JCMT observing runs. On the other hand,
we do have good values for the main beam efficiencies and so an
accurate conversion to main beam temperature (\tmb = \tas/$\eta_{\rm
MB}$) is possible. Therefore, we will display our spectra using the
main beam temperature scale. We report our fluxes in Table
\ref{masstable}.

\section{Discussion \label{disc}}

\subsection{Individual Galaxies}

{\em NGC 2273:} The NRAO 12-m detection of this galaxy is not strong.
Given the velocity width of the emission and the fact that all
the emission is located in the central few arcseconds \citep{pet02a} the large
beam of the 12-m dilutes the emission substantially.

We have JCMT CO \jto\ and \jtt\ spectra with strong detections
\citepalias[Figure \ref{morejcmtspectra}]{pet03} so despite the weak
emission in the NRAO spectrum for this region we expect to see
emission over the velocity range from 1600 to 2000 \kms. For
completeness we include the \twcott\ spectrum of this galaxy from
\citet{pet03} in Figure \ref{morejcmtspectra}.

{\em NGC 2859:} Despite our rather high sensitivity, this galaxy was
not detected in \twcoto\ with the NRAO 12-m, nor in \twcott\ with the
JCMT. A literature search shows that it was detected with the
3.5\arcmin\ beam of Arecibo and contains $2.1 \times 10^8$ \msun\ of
\ion{H}{1} \citep{bie77,war86}.

{\em NGC 2950:} Due to the similarity in LST with NGC 2859, we
were unable to attempt an NRAO 12-m observation of this galaxy.
It was observed with the JCMT but was not detected in \twcoto.
A literature search shows that \ion{H}{1} was not detected 
in the galaxy \citep{war86}.

{\em NGC 3081:}
We have JCMT CO \jto\ spectra for NGC
3081 (Figure \ref{morejcmtspectra})
that show emission over a region from 2200 to 2500 \kms, while in
the NRAO spectrum, we see no detectable line. This is likely due to
the high noise level in the NRAO spectra of NGC 3081 due to its low
declination.

{\em NGC 4340:} No emission was detected in \twcoto\ with the NRAO 12-m, so
followup observations were taken at the JCMT in \twcott\ with the
same result. This galaxy was not detected in \ion{H}{1} with
Arecibo \citep{gio83}.

{\em NGC 4371:} This galaxy is also not detected
with the NRAO 12-m in
\twcoto, JCMT in \twcott, or
Arecibo in \ion{H}{1} \citep{gio83}.

{\em NGC 4736:} This nearby galaxy was easily detected with the NRAO
12-m in \twcoto. The line profiles and peak strength for NGC 4736
agree well with those published in \citetalias{pet03}.

{\em NGC 5728:}
The emission in NGC 5728 is known to cover a wide range of velocities
from less than 2600 \kms\ to greater than 3050 \kms\ and is very
clumpy \citep{pet02a,sch88}. The line would nearly cover the entire
spectrometer which makes it difficult to determine the baseline for
the spectra of NGC 5728 shown in Figure \ref{nraospectra}. We have
used the very ends of the spectrometer to determine the baseline
level, and the result is a lumpy spectra with no strong noticeable
peaks, but a general tendency for the noise to remain slightly greater
than zero.  NGC 5728 is clearly detected in the JCMT spectrum shown in
Figure \ref{morejcmtspectra}.  

{\em NGC 5850:} This galaxy is clearly detected in the NRAO 12-m
spectrum.  There are no published single dish CO spectra for NGC 5850
so we cannot compare line profiles.  Single dish fluxes and
interferometric maps are published in \citet{leo00}, and they find
single dish gas masses comparable to ours, suggesting that our
pointing and calibration are correct.

{\em NGC 6951:} The NRAO 12-m line profile of NGC 6951 (Figure
\ref{nraospectra}) is single peaked which is noticeably different than
the JCMT \twcoto\ spectrum for this galaxy (see
\citetalias{pet03}). The profile of the NRAO spectrum more closely
resembles the CO \jtt\ spectrum taken with the JCMT at an offset of
(0\arcsec,$-$7\arcsec) as part of our 5-point mapping procedure
discussed in \citetalias{pet03}. In addition to this, the peak line
strength in the NRAO spectrum is much lower than the JCMT \twcoto\
spectrum, suggesting that pointing inaccuracies may have resulted in
pointing the telescope too far south, missing the strongest emission
in the northern part of the nucleus \citep{koh99} with the most
sensitive part of the beam.

We also point out here that after the inclusion of this galaxy
in our sample, more recent studies have suggested that it does not
contain a double bar as indicated by the earlier NIR surveys.
\citet{per00} find evidence that NGC 6951 may have contained
a nuclear bar at one point, but gas accumulation into the nucleus
may have resulted in its dissolution. We choose to keep it in
our sample since its presence does not affect any of our conclusions.

\subsection{Cumulative Results}

Despite our rather high sensitivity (\tmb(rms) $\approx$ 14 mK) we
have failed to detect CO \jto\ lines in NGC 2859, NGC 4340, or NGC
4371. We do not detect any galaxies that have not been previously
detected in the CO surveys of \citet{bra92}, \citet{mau99}, and
\citet{you95}. Looking at Table \ref{galtable} suggests that we are
predominantly detecting CO in later type galaxies, which is a result
known from previous studies \citep{you95}, but there is also an
apparent trend with nuclear activity.

Beside the galaxy names in Figures \ref{nraospectra} and
\ref{jcmtspectra} are codes (in parentheses) that indicate the types
of nuclear activity found in these galaxies. Seyfert 2s are marked as
``S2'' and LINERs are flagged with ``L'' \citep{ho97}. Galaxies
without any detected nuclear activity are not flagged and show no
signs of Seyfert, LINER, or starburst activity. All of
the galaxies that we have detected show signs of nuclear activity.  Of
the four galaxies that we have not detected in CO emission, none of
them show any signs of nuclear activity\footnote{\citet{ho97} gave NGC
2859 the uncertain classification of a ``transition object'', that is,
a galaxy showing signs of both an \ion{H}{2} and LINER
nucleus. However, the lines are weak and the spectra are noisy.}.   We
note that in performing a literature search, the galaxies that we have
not detected are quite a bit less studied than NGC 4736 and NGC 6951,
for example, and may harbor as yet undetected nuclear activity which
could change our small number statistics noticeably.

\subsection{Molecular Gas Mass \label{mass}}

The double barred galaxy models of \citet{fri93} and \citet{sha93}
suggest that there needs to be substantial amounts of molecular gas in
double barred galaxies. In fact, the molecular gas inflow in these
double barred galaxies may accumulate enough mass so that the nuclear
bar can become kinematically distinct \citep{fri93,pfe90}. Thus, we
may expect to see high molecular gas masses in the centers of these
double barred galaxies.

The intensity of the CO emission can be related to the molecular mass
using the equation \begin{equation} M_{\rm mol} = 1.61 \times
10^4~\left({\alpha\over{\alpha_{\rm Gal}}} \right)\left ({115~{\rm
GHz}\over{\nu}}\right )^2 ~ d^2_{\rm Mpc} ~ {S_{\rm CO} \over{R}} ~
\rm M_{\odot} \end{equation} \citep{wil90,wil95} where $S_{\rm CO}$ is
the \twcoto\ flux in Jy km s$^{-1}$, $R$ is the \twcoto/\joz\ line
ratio, $\nu$ is the frequency of the emission (230 GHz for the \jto\
transition), $d_{\rm Mpc}$ is the distance to the galaxy in Mpc,
$\alpha$ is the CO-to-H$_2$ conversion factor for that galaxy, and
$\alpha_{\rm Gal}$ is the Galactic value ($3\pm 1 \times 10^{20}$
cm$^{-2}$(K km s$^{-1}$)$^{-1}$, \citealt{str88}; \citealt{sco87}). We
use 24.7 Jy K$^{-1}$, 27.8 Jy K$^{-1}$ ($\eta_{\rm ap}$= 0.63, 0.56)
and 70.6 Jy K$^{-1}$ ($\eta_{\rm ap}$= 0.35) to convert our JCMT
(\jto, \jtt) and NRAO \jto\ data (respectively) from Kelvins ($T_{\rm
A}^\star$) to Janskys (\citealt{kra86}; JCMT Users Guide; NRAO 12-m
Users Manual). We assume a coupling efficiency ($\eta_{\rm c}$) of 0.7
to correct our observed fluxes to true fluxes. The CO-to-H$_2$
conversion factor ($\alpha$) is a globally averaged property of the
galaxy and hence there are uncertainties involved in its use in one
specific region of the galaxy and it is only accurate to within $\sim$
30\%. Our fluxes are typically accurate to about 10\%. The distances
for these relatively nearby galaxies are likely uncertain by at least
30\%. We therefore adopt a total uncertainty of 50\% in our mass
estimates.

For the galaxies where we have CO detections with other telescopes or
at other frequencies, we integrate over the velocity range where the
emission line was seen. For galaxies with no previous detections, we
integrate over a 400 \kms\ range centered on the rest velocity of the
galaxy (this region was excluded from the baseline subtraction). In
the cases where the integrated intensity is greater than the rms
noise, we give both the integrated intensity and the noise regardless
of how insignificant. We are not claiming these as detections, but are
simply using them as a more realistic value for the detection cut-off
limit.  If the integrated intensity is less than or equal to the rms
noise, the noise value is given as an upper limit. The results are
summarized in Table \ref{masstable}.

Table \ref{masstable} shows that there is a wide variety of molecular
gas masses in the inner regions of these galaxies. For the galaxies in
our sample that have also been detected with the JCMT, we find that
the masses determined here typically are lower or agree with the
masses determined with CO \jto\ data in \citetalias{pet03} within a
factor of two. The exceptions are NGC 6951 which is lower here by more
than a factor three and NGC 4736 which is higher by almost a factor of
two.  The discrepancy in NGC 6951 can likely be attributed to the
pointing offset discussed in \S\ref{disc}. The discrepancy in NGC 4736
can likely be attributed to the strong CO emission in the spiral arms
falling in the larger beam of the NRAO telescope \citep{won00}.  In
any case, the similarities between the masses obtained with the weaker
NRAO 12-m spectra (even in the cases where no obvious lines are
visible such as NGC 3081 and NGC 5728) and the masses obtained with
the JCMT \jto\ spectra gives us confidence that our mass estimates and
upper limits are accurate to at least a factor of two.

Of particular interest is the galaxy NGC 5850, whose spectrum
indicates that there is more than $10^9$ \msun\ of molecular gas in
the inner 29\arcsec.  This mass is comparable to the amount of
molecular gas in the entire Milky Way, but now contained in its inner
2.5 kpc radius \citep{dam93}. The optical size of this galaxy is
4\farcm 3$\times$3\farcm 7 ($D_{25} \times d_{25}$), which corresponds
to 43$\times$37 kpc at its distance of 34 Mpc. This clearly makes it
the largest galaxy in our sample (the second runner up is NGC 5728 at
33$\times$19 kpc). Given its rather strong primary bar, it is possible
that the large quantity of gas in the inner regions of this galaxy may
have been transported inward by the inflow mechanisms known to be
associated with bar perturbations. The high resolution CO maps of this
galaxy \citep{leo00} detect only $6.7 \times 10^7$ \msun\ of molecular
gas, mostly concentrated in a small off-center peak of emission
approximately 8\arcsec\ north of the galactic center. On the other
hand, their single dish IRAM 30-m CO \joz\ map of the entire primary
bar detects $3.4 \times 10^9$ \msun.  \citet{leo00} point out that
this galaxy is surprisingly quiescent given the large amounts of
molecular gas, and propose that the reason for this is that the
molecular gas is below the critical surface density for gravitational
instabilities \citep{ken89}.

The large size of NGC 5850 and the fact that it is the only quiescent
galaxy with a strong detection lead us to wonder if we are detecting
emission lines in predominantly the largest galaxies (with possibly
the largest molecular gas reservoirs). All the other galaxies are in
the 15 to 20 kpc (major axis) size range with the exception of NGC
2859, NGC 5728, and NGC 5850 who have major axes of 28, 33, and 43 kpc
respectively. The strongest line occurs in the closest galaxy, NGC
4736, which is incidentally one of the smallest in our sample with
major axis of $\sim$14 kpc. So it seems that we are not detecting CO
emission preferentially in larger galaxies. Since NGC 5850 is the
second most distant galaxy in our sample, it also appears that we are
not preferentially detecting emission from the closest galaxies.

Since we are searching for emission with the CO \jto\ line, our lack
of success in finding bright candidates may not be the result of a
lack of molecular gas, but that the gas in these galaxies is very cool
and possibly at a low density. Is it possible that all of the
molecular emission is dominated by \joz\ emission and it is not
excited into the \jto\ levels enough to be detected? In our mass
calculation, we assume a \jto/\joz\ line ratio of 0.7. In the Local
Thermodynamic Equilibrium approximation, in order to achieve this line
ratio, the gas must be at a temperature of only 7 K. This temperature
is low enough that it can be maintained by cosmic ray heating
\citep{gol78}. Higher values of the \jto/\joz\ line ratio will act to
decrease our molecular gas mass, meaning that the mass values quoted
here are likely upper limits.

Another possible explanation for the low molecular gas mass may be
that the gas is just not located in the inner 29\arcsec. There are
observations of other galaxies that contain large molecular rings that
seem to have prevented any of the molecular gas from reaching the
nucleus \citep[\eg\ NGC 7331;][]{she00}. We will need spectra covering
a wider field of view to verify if this is happening in any of these
galaxies.

\subsection{Implications for Double Barred Galaxy Models \label{models}}

All the galaxies in this survey 
are known to contain the NIR isophote twists believed to be the
signature of a double barred galaxy. As mentioned in \S\ref{intro},
the models of double barred galaxies by \citet{sha93} and
\citet{fri93} require large amounts of molecular gas in the nuclear
regions of these galaxies in order to sustain long-lived double
bars. The gas requirements vary from 4 to 6\% globally
\citep{sha93,fri93} to as little as 2\% in the nucleus
\citep{fri93}. This gas is required to provide the dissipation needed
to prevent the stellar component from dynamically heating so much that
the nuclear bar is destroyed. We point out that we are studying the
molecular gas properties in {\it only the nuclear region} so we expect
to see gas mass fractions on the order of 2\% or more.

The molecular gas content of the galaxies for which we have detections
are in agreement with the model requirements and are discussed
elsewhere \citep{pet02a,leo00,sak99}. We will focus our attention here
on the galaxies where we have failed to detect a strong molecular gas
component. These are the galaxies that are difficult to explain in
light of the current models of double barred galaxies.

As mentioned in \S\ref{disc}, the galaxies where we have not detected
any CO emission are in general less well studied. As such, we are not
able to find direct measurements of the masses of all the galaxies in
our sample in the literature. We have estimated the mass for these
galaxies from their blue magnitude ($m_{B(T)}$ in
\citealt{rc3}). Table \ref{starmasstable} shows the apparent blue
magnitudes for our sample of galaxies, the absolute magnitude, and the
luminosity in solar luminosities.  Independent measurements for the
masses of some galaxies in our sample exists in the
literature. \citet{rub80} determined the dynamical mass of the disk
(the dominant source of the blue light) of NGC 5728 to be $\sim 8
\times 10^{10}$ \msun; \citet{mar93} find the mass of NGC 6951 to be
$1.3 \times 10^{11}$ \msun; \citet{smi91} determine the mass of NGC
4736 to be $4 \times 10^{10}$ \msun. These masses are in acceptable
agreement with our estimates considering the uncertainties associated
with our technique.  It seems, though, that our method is
underestimating the galaxy mass by a factor of 2 to 4, which is likely
the result of the blue light missing much of the older, redder stellar
population of stars in these galaxies. Our estimate therefore provides
a lower limit to the true galaxy mass and thus an upper limit to the
gas mass fraction.

The limits on the molecular gas masses for the undetected galaxies in
our sample range from as high as $3 \times 10^7$ \msun (for NGC 2859)
down to $7 \times 10^6$ \msun\ for NGC 4371. Using the galaxy masses
shown in Table \ref{starmasstable}, this corresponds to a molecular
gas mass fraction of 0.07\% and 0.05\% for NGC 2859 and NGC 4371,
respectively.  These low gas mass fractions suggest that some of these
galaxies do not contain enough molecular gas to be able to support
the nuclear bars observed in the models of \citet{fri93} and
\cite{sha93}.

There are four possible explanations that can bypass this potential
problem. The first possibility is that the molecular gas is not
confined to the nuclei of these galaxies. A circumnuclear CO
morphology is seen in other galaxies \citep[\eg, NGC 7331;][]{she00},
so it is possible that the molecular gas is outside of the inner
29\arcsec\ covered by the NRAO beam for the seemingly gas deficient
galaxies in our sample. We will need CO observations over a larger
area in order to determine if this is the case, but the models of
\citet{fri93} still require that the inner kpc contain at least 2\%
gas. If we assume similar disk and bulge profiles for NGC 2859 as
those adopted for NGC 5728 by \citet{rub80}, we estimate that roughly
one tenth of the stellar mass (bulge + disk) is contained in the inner
29\arcsec\ of NGC 2859. This translates into a gas to mass ratio of
{\it 0.7\% for the inner 3 kpc} of NGC 2859, so the lack of CO in this
galaxy is still a problem for the \citet{fri93} model. Even if the gas
is located in a large ring beyond our NRAO 12-m beam, it will be part of
the main disk and will not have much of an
impact on the cooling of the nuclear regions. The failure of the
large beams of the \ion{H}{1} studies to detect any gas in NGC 2950,
NGC 4340, or NGC 4371 suggests that the entire disks of these galaxies
are very gas poor.

Another possibility is that the gas in our galaxies is not in
molecular form. The gaseous components of the models are basically a
dissipation mechanism that acts to prevent the stellar components from
being dynamically heated. Models of these galaxies generally treat
this gas as being primarily molecular, contained in regions of high
density and low filling factor \citep[\eg,][]{com85}. Molecular gas
also has a higher cooling capacity, since it contains many more
emission lines available to it compared to atomic gas. It would take
much more \ion{H}{1} gas to dissipate as much energy as molecular gas.
We have searched the literature for \ion{H}{1} observations of these
four galaxies that are undetected in CO. Only NGC 2859 has been
detected in \ion{H}{1} by various authors
\citep{bie77,gio83,war86,esk91}. An \ion{H}{1} mass of $2 \times 10^8$
\msun\ was determined by \citet{bie77} for the entire disk of NGC
2859, which corresponds to a gas mass ratio of 0.5\% {\it globally},
which is still less that required by the models of \citet{fri93} and
\citet{sha93}. Additionally, 60 $\mu$m and 100 $\mu$m fluxes suggest
global star formation rates less than 0.1 \msun/year
\citep{esk91,ken98}, supporting the claim that these galaxies are gas
deficient.

The third possibility is that the NIR isophote twists are not
correlated with the gas properties at all. It is possible that the NIR
isophote twists are caused by a triaxial stellar bulge, as originally
proposed by \citet{kor79}. In this scenario, the lack of molecular gas
is not a problem, because there is not much molecular gas in the
bulges of galaxies anyway. Evidence against the triaxial bulge model
is discussed in \citetalias{pet02a} where the existence of a nuclear
molecular feature that aligns with the isophote twists supports the
existence of a true nuclear bar in the disk of NGC 2273. In addition,
detailed analysis of the NIR isophotes indicates that the variations
in position angle and bar ellipticity observed in some galaxies cannot
be the result of triaxial stellar bulges, but must be produced by
nuclear stellar bars \citep{jun97}. Observations of deprojected
nuclear bar/primary bar offset angles have also ruled out the
possibility that the isophote twists are the result of collections of
stars trapped in $x_2$ orbits \citep{fri93}.

Finally, if the NIR isophote twists are caused by nuclear stellar
bars, but we do not see a significant gas mass fraction in the
nucleus, our observations thus favor models that can produce nuclear
stellar bars without a molecular gas counterpart.  Recently,
\citet{mac00} have discovered a small family of orbits that are
capable of sustaining nuclear stellar bars. In light of the lack of gas
in some of the galaxies discussed here, we believe that of all models
so far, this is the most promising explanation for these instances of
double barred galaxies.

It is possible that different mechanisms are at work in different
galaxies, and they need to be studied on a case by case basis to
determine if the isophote twists are the result of a nuclear bar or a
triaxial stellar bulge. In either case, we will need either more
sensitive arrays or a sub-millimeter interferometer in the southern
hemisphere in order to obtain high resolution CO maps for a larger
number of these galaxies. Additionally, high resolution studies
of the star formation histories of these galaxies will help determine
which stage these galaxies occupy in the evolution of
double-barred galaxies.

\section{Summary \label{conc}}

In an attempt to find double barred galaxies that are bright in CO
emission, we have obtained \twcoto\ spectra for nine galaxies with the
NRAO 12-m Telescope. In the cases where no emission was found with the
NRAO 12-m, we obtained higher resolution \twcott\ and \jto\ spectra
with the James Clerk Maxwell Telescope. There is only one additional
detection in the JCMT spectra of these galaxies despite reaching
sensitivities of 4 mK (\tas). We detect emission in five of these
galaxies. All five galaxies detected exhibit some form of nuclear
activity, while the galaxies that were not detected are
quiescent and show no signs of any nuclear activity.  Thus, within our
small sample, the CO emission seems to be detected predominantly in
galaxies that harbor some form of nuclear activity (\eg\ Seyfert,
LINER). We note that the quiescent galaxies are less well studied than
the active galaxies in our sample, so it may be that they harbor some
form of nuclear activity that has yet to be discovered.

Some models of double barred galaxies suggest that they should be gas
rich in order provide a means of dissipating energy that would
otherwise heat the stellar population and subsequently destroy the
nuclear bars.  We use the CO fluxes to estimate the amount of
molecular gas in the centers of these galaxies and we find gas masses
that range from $7 \times 10^6$ \msun\ to more than $\sim 10^9$ \msun.

The lack of CO and \ion{H}{1} detections places very strict limits on
the amounts of gas in these galaxies. For some galaxies, there must be
less than a few $\times 10^6$ \msun\ of molecular gas, which (assuming
these galaxies are typical disk galaxies) corresponds to gas mass
fractions of 0.05 to 0.8\% (depending on the assumed mass
distribution). These very low gas mass fractions suggest that, contrary
to many models, {\em large amounts of molecular gas are not required}
to sustain double-bars in the nuclei of some galaxies.

\acknowledgments 

G.~R.~P.~is supported by NSF grant AST 99-81289 and by the State of
Maryland via support of the Laboratory for Millimeter-Wave
Astronomy. This research has also been supported by a research grant
to C.~D.~W.~from NSERC (Canada). This research has made use of the
NASA/IPAC Extragalactic Database (NED) which is operated by the Jet
Propulsion Laboratory, California Institute of Technology, under
contract with the National Aeronautics and Space Administration.  We
wish to thank the anonymous referee for helpful comments that greatly
improved the quality of this paper.  We also wish to thank Rob Ivison,
Susie Scott, Tracy Webb, and the staff of the JCMT for their help with
the observations taken remotely during non-cosmology weather. GRP
wishes to thank the extremely helpful staff at the NRAO 12-m for all
their assistance in the remote observing run in February 2000, which
went more smoothly than many observing runs for which he himself was
present.

\clearpage

\figcaption[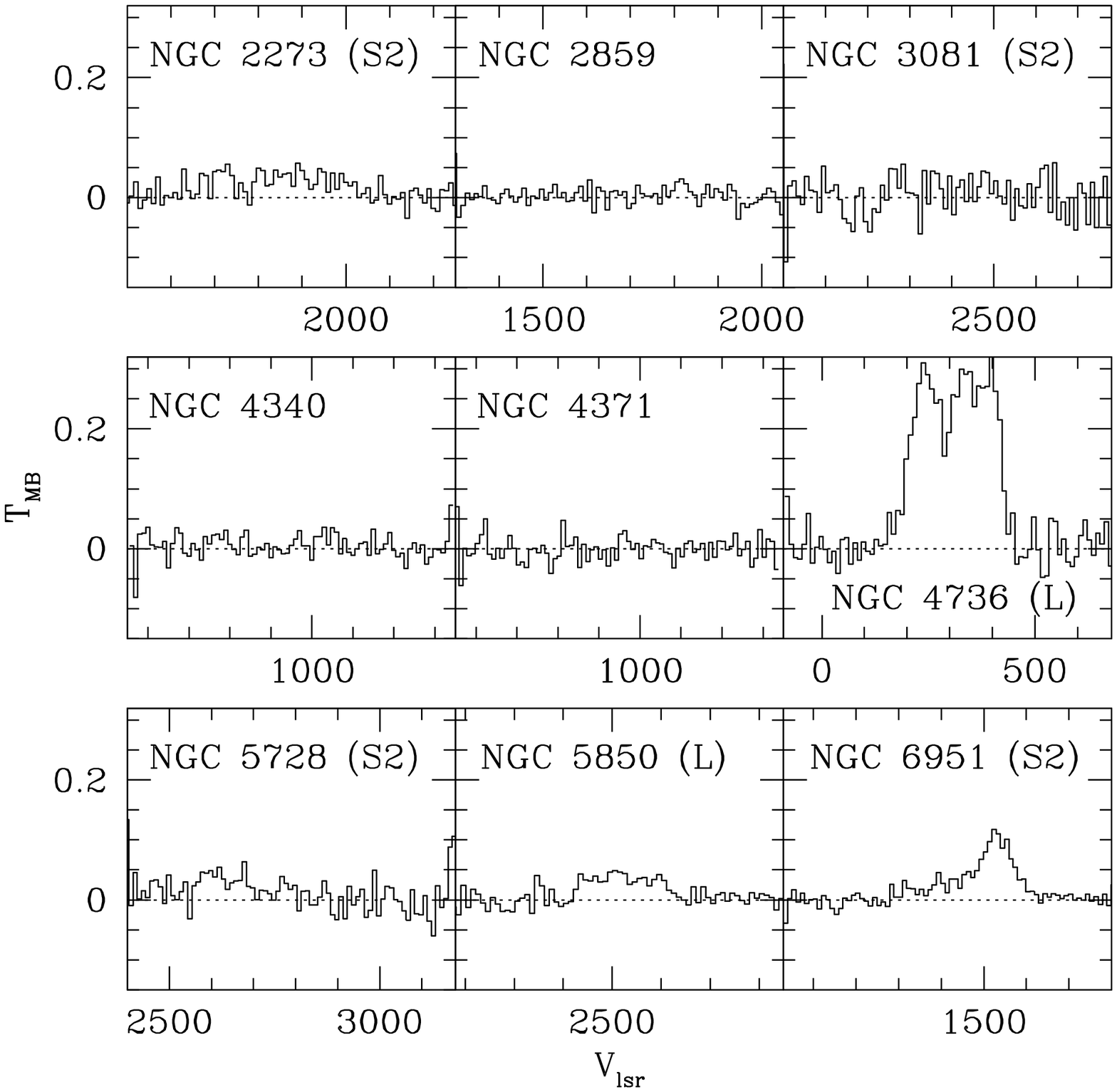]{\twcoto\ spectra taken at the NRAO 12-m of a
sample of galaxies with NIR isophote twists and thought to contain
double bars. The spectra cover the inner 29\arcsec\ regions of the galaxy
nuclei, which are predicted to be gas rich by the models of
\citet{sha93} and \citet{fri93}. The type of nuclear activity exhibited
is shown after the galaxy name (S2 = Seyfert 2; L = LINER).  Note that
we detect CO emission primarily from galaxies with some form of nuclear
activity. The large tick marks on the velocity axis correspond to 500
\kms\ intervals, while the smaller tick marks are every 100 \kms. Higher
recession velocities are to the right.
\label{nraospectra}
}

\figcaption[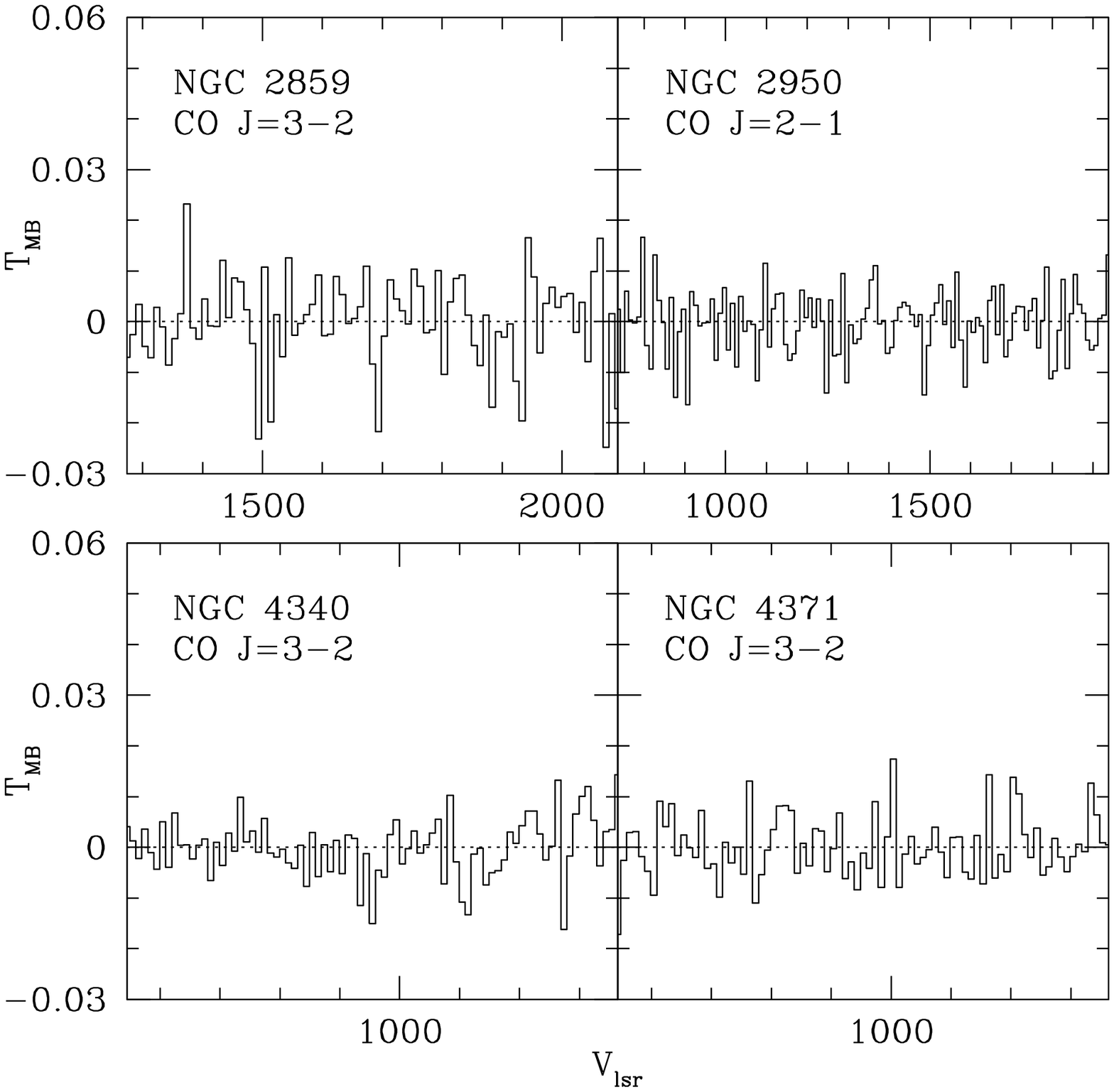]{\twcoto\ and \twcott\ spectra taken at the JCMT of
a sample of galaxies thought to contain double bars. The \jto\ spectra
cover the inner 21\arcsec\ while the \jtt\ spectra cover 14\arcsec.
Note that despite the rather low noise, we do not detect any emission from
these four galaxies.
\label{jcmtspectra}
}

\figcaption[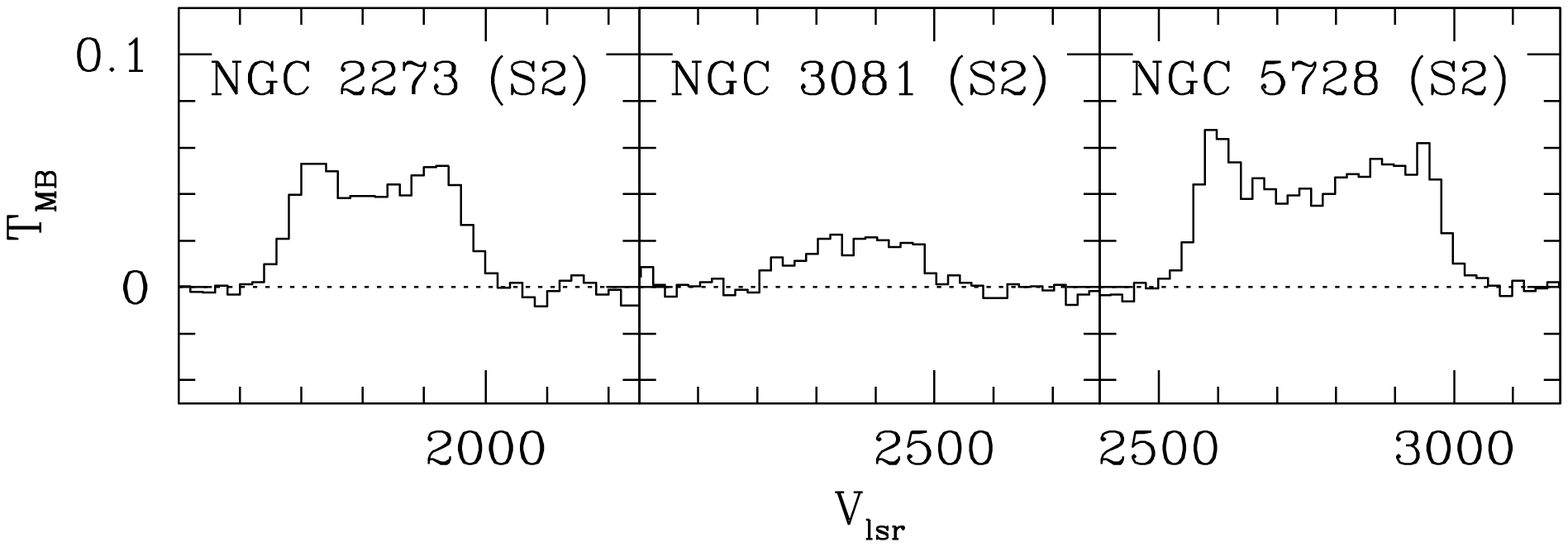]{JCMT \twcoto\ spectra from Paper II of three
galaxies where the \twcoto\ detections with the NRAO were not convincing
(see Figure \ref{nraospectra}). 
\label{morejcmtspectra}
}

\begin{deluxetable}{lccccccc}
\tablecolumns{8}
\tablewidth{0pc}
\tablecaption{Galaxy Parameters \label{galtable}}
\tablehead{
\colhead{Galaxy} & \colhead{$\alpha$} & \colhead{$\delta$} & \colhead{$V_{\rm lsr}$} & \colhead{D$_{75}$} & \colhead{   scale    } & \colhead{RC3} & \colhead{companion?}   \\
\colhead{~     } & \colhead{ (2000) } & \colhead{ (2000) } & \colhead{  (\kms)     } & \colhead{ (Mpc)  } & \colhead{(pc/\arcsec)} & \colhead{~  } & \colhead{    }   \\
          } 
\startdata
NGC 2273  &  6\h50\m09\fs8 & $+$60\arcdeg50\arcmin49\arcsec & 1870 & 25 & 120 & SB(r)a           & NGC 2273B  \\
NGC 2859  &  9\h24\m19\fs5 & $+$34\arcdeg30\arcmin43\arcsec & 1687 & 22 & 110 & (R)SB(r)0        & no \\
NGC 2950  &  9\h42\m35\fs1 & $+$58\arcdeg51\arcmin05\arcsec & 1337 & 18 &  87 & (R)SB(r)0        & no \\
NGC 3081  &  9\h59\m30\fs6 & $-$22\arcdeg49\arcmin41\arcsec & 2385 & 32 & 160 & (R$_1$)SAB(r)0/a & no \\
NGC 4340  & 12\h23\m35\fs8 & $+$16\arcdeg43\arcmin16\arcsec &  950 & 13 &  63 & SB(r)0           & NGC 4350? \\
NGC 4371  & 12\h24\m55\fs5 & $+$11\arcdeg42\arcmin10\arcsec &  943 & 13 &  63 & SB(r)0           & no \\
NGC 4736  & 12\h50\m53\fs4 & $+$41\arcdeg07\arcmin02\arcsec &  308 &  4 &  19 & (R)SA(r)ab       & no      \\
NGC 5728  & 14\h42\m23\fs8 & $-$17\arcdeg15\arcmin03\arcsec & 2788 & 37 & 180 & (R$_1$)SAB(r)a   & no \\
NGC 5850  & 15\h07\m07\fs5 &  $+$1\arcdeg32\arcmin43\arcsec & 2556 & 34 & 160 & SB(r)b           & NGC 5846 \\
NGC 6951  & 20\h37\m11\fs6 & $+$66\arcdeg06\arcmin12\arcsec & 1424 & 19 &  92 & SAB(rs)bc        & no     \\
\enddata
\tablecomments{All distances are taken from the NASA-IPAC Extragalactic
Database (NED) and assume a Hubble Constant of 75 \kms\ (Mpc)$^{-1}$.}
\end{deluxetable}

\begin{deluxetable}{lccccc}
\tablecolumns{6}
\tablewidth{0pc}
\tablecaption{Observing Parameters \label{obstable}}
\tablehead{
\colhead{Galaxy} & \colhead{Line} & \colhead{Tele.} & \colhead{t$_{\rm int.}$} & \colhead{T$_{\rm sys}$} & \colhead{r.m.s.\tablenotemark{a}}  \\
\colhead{~} & \colhead{(\twco)} & \colhead{~} &  \colhead{(h:m)} & \colhead{(K)} & \colhead{(mK \tmb)}  \\
          } 
\startdata
NGC 2273   & \jto & NRAO & 1:46 &  516 & 16 \\
\nodata    & \jto & JCMT & 0:30 &  322 & 8  \\
NGC 2859   & \jto & NRAO & 1:58 &  467 & 14 \\
\nodata    & \jtt & JCMT & 2:20 &  630 & 6  \\
NGC 2950   & \jto & JCMT & 0:30 &  254 & 6  \\
NGC 3081   & \jto & NRAO & 2:09 & 1012 & 28 \\
\nodata    & \jto & JCMT & 2:20  & 463 & 7  \\
NGC 4340   & \jto & NRAO & 1:58 &  199 & 13 \\
\nodata    & \jtt & JCMT & 2:00 &  485 & 5  \\
NGC 4371   & \jto & NRAO & 1:58 &  511 & 15 \\
\nodata    & \jtt & JCMT & 2:00 &  503 & 5  \\
NGC 4736   & \jto & NRAO & 0:35 &  468 & 25 \\
NGC 5728   & \jto & NRAO & 1:58 &  605 & 18 \\
\nodata    & \jto & JCMT & 0:50 &  467 & 9  \\
NGC 5850   & \jto & NRAO & 1:58 &  453 & 13 \\
NGC 6951   & \jto & NRAO & 3:20 &  424 & 10 \\
\enddata
\tablenotetext{a}{Given for a velocity resolution of 10 \kms\ which corresponds to 7.68 MHz at 230 GHz and 11.5 MHz at 345 GHz.}
\end{deluxetable}

\begin{deluxetable}{lccccc}
\tablecolumns{6}
\tablewidth{0pc}
\tablecaption{CO Fluxes and Molecular Gas Masses \label{masstable}}
\tablehead{
\colhead{Galaxy} & \colhead{V limits} & \colhead{Transition (tele.)}  & \colhead{Flux\tablenotemark{a}}&  \colhead{$r_{\rm int}$\tablenotemark{b}} & \colhead{Gas Mass}  \\

\colhead{~} & \colhead{(\kms)} &\colhead{(\twco)} & \colhead{(K \kms\ \tmb)}& \colhead{(kpc)} & \colhead{(\msun)}  \\
          } 
\startdata
NGC 2273   & 1600-2020 & \jto\ (NRAO)&   9.8 $\pm$ 1.0 & 1.7 & $  8.2 \times 10^8$  \\
\nodata    & \nodata   & \jto\ (JCMT)&   19.7$\pm$ 0.5 & 1.3 &  $1.3 \times 10^9$   \\
NGC 2859   & 1490-1890 & \jto\ (NRAO)&  (2.6 $\pm$ 0.9)& 1.6 & $ (2.0 \times 10^8)$  \\
\nodata    & \nodata   & \jtt\ (JCMT)&   $<$ 0.40      & 0.8 & $< 2.9 \times 10^7$  \\
NGC 2950   & 1140-1540 & \jto\ (JCMT)&   $<$ 0.33      & 0.9 & $< 1.5 \times 10^7$  \\
NGC 3081   & 2200-2550 & \jto\ (NRAO)& (3.6 $\pm$ 1.8) & 2.3 & $ (6.2 \times 10^8)$  \\
\nodata    & \nodata   & \jto\ (JCMT)&  4.6 $\pm$ 0.3  & 1.7 & $  6.5 \times 10^8$   \\
NGC 4340   &  750-1150 & \jto\ (NRAO)& (2.8 $\pm$ 0.8) & 0.9 & $ (8.1 \times 10^7)$  \\
\nodata    &  \nodata  & \jtt\ (JCMT)&   $<$ 0.32      & 0.4 & $< 7.7 \times 10^6$  \\
NGC 4371   &  740-1140 & \jto\ (NRAO)&   $<$ 1.0       & 0.9 & $< 2.7 \times 10^7$  \\
\nodata    &  \nodata  & \jtt\ (JCMT)&   $<$ 0.29      & 0.4 & $< 7.0 \times 10^6$  \\
NGC 4736   &  100-450  & \jto\ (NRAO)& 61.7 $\pm$ 1.4  & 0.3 & $  1.7 \times 10^8$  \\
NGC 5728   & 2500-3050 & \jto\ (NRAO)& 7.4 $\pm$ 1.3   & 2.6 & $  6.1 \times 10^8$  \\
\nodata    & \nodata   & \jto\ (JCMT)& 24.6 $\pm$ 0.6  & 1.9 &  $  1.7 \times 10^9$  \\
NGC 5850   & 2400-2650 & \jto\ (NRAO)& 6.9 $\pm$ 0.7   & 2.3 & $  1.3 \times 10^9$  \\
NGC 6951\tablenotemark{c}  & 1250-1620 & \jto\ (NRAO)& 13.4 $\pm$ 0.6  & 1.3 & $  9.7 \times 10^8$  \\
\enddata
\tablecomments{
For calculating gas masses for NGC 2273, NGC 5728, NGC 6951 we adopt CO
\jto/\joz\ ratios of 0.88, 1.96, and 0.59 respectively \citep{pet03}.
For the other galaxies we have assumed a \twcoto/\joz\ ratio of 0.7 and
(where necessary) a \jtt/\jto\ line ratio of 1, similar to the values
found for other double barred galaxies \citep{pet03}. The flux values for
the questionable detections (where the integrated intensities are
larger than three times the noise associated with that region) are enclosed in parentheses.
}
\tablenotetext{a}{The beam size for the NRAO 12-m at CO \jto\ is 29\arcsec. The
beam size for the JCMT is 21\arcsec\ at CO \jto\ and 14\arcsec\ at \jtt.}
\tablenotetext{b}{The radius interior to which the flux/mass has been measured.}
\tablenotetext{c}{The pointing may have been off; see text for details.}
\end{deluxetable}

\begin{deluxetable}{lcccc}
\tablecolumns{5}
\tablewidth{0pc}
\tablecaption{Galaxy Masses Estimated from Blue Light \label{starmasstable}}
\tablehead{
\colhead{Galaxy} & \colhead{$m_{B(T)}$} & \colhead{$M_{B}$} & \colhead{$L_B$}& \colhead{Mass} \\
\colhead{~} & \colhead{(app.)} & \colhead{(abs.)} & \colhead{(L$_\odot$)}& \colhead{(\msun)}
          } 
\startdata
NGC 2273  & 12.55 & $-$19.44 & 8.7 $\times 10^9$   &  2.6 $\times 10^{10}$  \\
NGC 2859  & 11.83 & $-$19.87 & 1.3 $\times 10^{10}$&  3.9 $\times 10^{10}$  \\
NGC 2950  & 11.84 & $-$19.44 & 8.7 $\times 10^9$   &  2.9 $\times 10^{10}$  \\
NGC 3081  & 12.85 & $-$19.68 & 1.1 $\times 10^{10}$&  3.3 $\times 10^{10}$ \\
NGC 4340  & 12.10 & $-$18.47 & 3.6 $\times 10^9$   &  1.1 $\times 10^{10}$  \\
NGC 4371  & 11.79 & $-$18.78 & 4.7 $\times 10^9$   &  1.4 $\times 10^{10}$  \\
NGC 4736  &  8.99 & $-$19.02 & 5.9 $\times 10^9$   &  1.8 $\times 10^{10}$  \\
NGC 5728  & 12.57 & $-$20.27 & 1.9 $\times 10^{10}$&  5.7 $\times 10^{10}$  \\
NGC 5850  & 11.54 & $-$21.18 & 4.3 $\times 10^{10}$&  1.3 $\times 10^{11}$  \\
NGC 6951  & 11.64 & $-$19.75 & 1.2 $\times 10^{10}$&  3.6 $\times 10^{10}$  \\
\enddata
\tablecomments{$m_{B(T)}$ is the apparent blue magnitude of the galaxy
extrapolated to infinite radius (RC3). We adopt +5.41 for the Suns
absolute blue magnitude \citep{all64} and note that a change of
$\sim$0.3 in the galactic magnitude would result in a factor
of $\sim$2 variation in the luminosity. The last column assumes a mass
to light ratio of 3, which is typical for barred spiral galaxies
\citep{for92}.
}
\end{deluxetable}

\clearpage

\plotone{f1.eps}
\vfil
\plotone{f2.eps}
\vfil
\plotone{f3.eps}

\end{document}